\documentclass{article}    

\usepackage{graphicx}
\usepackage{dcolumn}
\usepackage{bm}

\begin{document}           

\title{Extending the redshift-distance relation in Cosmological General Relativity to higher redshifts}  
\author{
\textbf{John G. Hartnett}\\
School of Physics, the University of Western Australia,\\
 35 Stirling Hwy, Crawley 6009 WA Australia\\
\textit{john@physics.uwa.edu.au}}

\maketitle  

\begin{abstract}
The redshift-distance modulus relation, the Hubble Diagram, derived from Cosmological General Relativity has been extended to arbitrarily large redshifts. Numerical methods were employed and a density function was found that results in a valid solution of the field equations at all redshifts. The extension has been compared to 302 type Ia supernova data as well as to 69 Gamma-ray burst data. The latter however do not not truly represent a `standard candle' as the derived distance modulii are not independent of the cosmology used. Nevertheless the analysis shows a good fit can be achieved without the need to assume the existence of dark matter.

The Carmelian theory is also shown to describe a universe that is always spatially flat. This results from the underlying assumption of the energy density of a cosmological constant $\Omega_{\Lambda} = 1$, the result of vacuum energy.  The curvature of the universe is described by a \textit{spacevelocity} metric where the energy content of the curvature at any epoch is $\Omega_K = \Omega_{\Lambda} - \Omega = 1-\Omega$, where $\Omega$ is the matter density of the universe. Hence the total density is always $\Omega_K + \Omega = 1$. 
\end{abstract}


\section{\label{sec:Intro} Introduction}
  Carmeli's cosmology, also referred to as Cosmological General Relativity (CGR), is a space-velocity theory of the expanding universe. It is a description of the universe at a particular fixed epoch of cosmic time $t$ and involves a new dimension $v$, the velocity of the expansion of the fabric of space itself, in which the galaxies are sitting.  For a full introduction to the theory, the reader should become familiar with either of the late Moshe Carmeli's books \textit{Cosmological Special Relativity, 2nd Edition} \cite{Carmeli2002a} or \textit{Cosmological Relativity} \cite{Carmeli2006} .
  
   In the 1920s Edwin Hubble discovered that the observed redshift $z$ in the light emitted from a distant source is directly proportional to the distance $D$ to the source, viz. $v = H_{0}\,D$, where $H_{0}$ is Hubble's constant.  Carmeli incorporated this law as a fundamental axiom, with $H_0^{-1} \approx \tau$, a universal time constant, into a general 4D Riemannian geometrical theory satisfying the Einstein field equations (Ref. \cite{Carmeli2002a}, appendix A).  The observables, redshift and distance (Hubble distance), are the coordinates of his theory. 
   
  The expanding universe is then described by its \textit{spacevelocity}  ``phase space''  coordinates $(v,r,\theta,\phi)$, which do not involve time $t$ but this new dimension $v$. Cosmic time is measured from the present epoch, with $t = 0$, back toward the beginning with $t = \tau$, but determined from the expansion velocity of the fabric of space, i.e $t \rightarrow \tau$ as $v \rightarrow c$. In this case the galaxies act as tracers embedded in the expanding space. This is different to the approach in the 4D Friedmann theory with coordinates $(t,r,\theta,\phi)$. But Carmeli never solved the full 5D theory involving the time dimension as well. In section \ref{sec:flatspace} I show how the time dimension is included in a general 5D CGR theory for the universe. Surprisingly, or may be not, what results is that the universe is Euclidean at all epochs. 
  
  Usually the Hubble law is stated in terms of the velocities of the receding galaxies, when actually those velocities are determined from the measured redshifts of spectral lines in the galaxy light. Carmeli used this fact to formulate his new theory. In practice redshift, and not velocity, is what astronomers actually measure, as well as distance determined from the luminosity of the source. In section \ref{sec:Comparison} I describe how the CGR theory can be compared with observations using luminosity distance as an independent variable, and how this is different to FRW theory. In section \ref{sec:Extended} I discuss how to extend the theory to a much greater range of validity in redshift space by solving Carmeli's equation directly (though numerically) for redshifts, instead of working in velocity space. In section \ref{sec:Fits} I show how well the theory fits the observations, including both type Ia supernova and Gamma-ray bursts.

\section{\label{sec:phasespace}Phase space equation}
  
Carmeli initially developed his \textit{special} theory from considerations of a universe without matter, then later solved the field equations for a fully covariant \textit{general} 4D theory. He noticed the analogue with special relativity. In Cosmological Special Relativity (see chap.2 of \cite{Carmeli2002a}), instead of the Minkowski line element of special relativity we can write the line element as
\begin{equation} \label{eqn:CSRlineelement}
ds^2= \tau^{2}dv^{2}-dr^{2},
\end{equation}
where $dr^2=dx^2+dy^2+dz^2$. Now by equating $ds=0$ it follows from (\ref{eqn:CSRlineelement}) that $\tau dv = dr$ assuming the positive sign for an expanding universe. This is then the Hubble law in the small $v$ limit, where $\tau$ is the Hubble-Carmeli time constant, the inverse of the Hubble constant in the limit of weak gravity. 

One sees the similarity, where instead of the light cone of special relativity, we get the concept of a velocity cone. Galaxies are distributed in the universe such that as their recession speed approaches the speed of the expansion of space they approach the limits of the cone. So, in general, this theory requires that the expansion of the universe be described by $ds =0$. The following describes the result when matter is considered in the general theory.

The line element in CGR \cite{Carmeli2002b}
\begin{equation} \label{eqn:lineelement}
ds^2= \tau^{2}dv^{2}-e^{\xi}dr^{2}-R^2(d\theta^2+ sin^2\theta d\phi^2),
\end{equation}
represents a spherically symmetrical isotropic universe, that is not necessarily homogeneous. 

Using spherical coordinates ($r,\theta, \phi$) and the isotropy condition $d\theta = d\phi = 0 $ in (\ref{eqn:lineelement}) then $dr$ represents the radial co-ordinate distance to the source and it follows from (\ref{eqn:lineelement}) that
\begin{equation} \label{eqn:4Dmetric}
\tau^{2}dv^{2}-e^{\xi}dr^{2}=0,
\end{equation}
where $\xi$ is a function of $v$ and $r$ alone. This results in
\begin{equation} \label{eqn:4Dmetricderiv}
\frac{dr} {dv} = \tau e^{-\xi/2},
\end{equation}
where the positive sign has been chosen for an expanding universe.

Carmeli found a solution to his field equations, modified from Einstein's, (see \cite{Hartnett2005} and \cite{Carmeli2002a,Carmeli2002b}) which is of the form
\begin{equation} \label{eqn:soln1field2}
e^{\xi}= \frac{{R'^2}}{1 + f(r)},
\end{equation}
with $R = r$ and hence $R' = 1$, which must be positive. From the field equations and (\ref{eqn:soln1field2}) we get a differential equation
\begin{equation} \label{eqn:soln1field3}
f' + \frac{f}{r} = - \kappa  \tau^{2} \rho_{eff} r,
\end{equation}
where $f(r)$ is function of $r$ and satisfies the condition $f(r) + 1 > 0$. The prime is the derivative with respect to $r$. Here $\kappa = 8 \pi G /c^{2} \tau^{2}$ and $\rho_{eff} = \rho -\rho_{c}$ where $\rho$  is the averaged matter density of the Universe and $\rho_{c}=3/8 \pi G \tau^{2}$ is the critical density. Finally the solution to the inhomogeneous equation (\ref{eqn:soln1field3}) was found to be 
\begin{equation} \label{fr}
f(r)= -\frac{\kappa}{3}  \tau^{2} \rho_{eff} r^{2} = (1-\Omega) \frac{r^2}{c^2\tau^2}.    
\end{equation} 
Therefore this requires that
\begin{equation} \label{eqn:condition}
1 + (1-\Omega) \frac{r^2}{c^2\tau^2} > 0.
\end{equation}

In the above Carmelian theory it initially assumed that the Universe has expanded over time and at any given epoch it has an averaged density $\rho$, and hence $\rho_{eff}$. The solution of the field equations has been sought on this basis. However because the Carmeli metric is solved in an instant of time (on a cosmological scale) any time dependence is neglected. In fact, the general time dependent solution has not yet been found. But since we observe the expanding Universe with the coordinates of Hubble at each epoch (or redshift $z$) we see the Universe with a different density $\rho(z)$ and an effective density  $\rho_{eff} (z)$. In section \ref{sec:Extended} this fact is taken into account to extend the validity of the solution.

\section{\label{sec:Comparison}Comparison with observation}

In order to compare CGR theory with observations one requires a relationship that gives proper distance as a function of redshift. From (\ref{eqn:4Dmetricderiv}), (\ref{eqn:soln1field2}) and (\ref{fr}) we get the differential equation 
\begin{equation} \label{eqn:phasespacederiv}
\frac{dr}{dv}= \tau \sqrt{1+ \left(\frac{1-\Omega}{c^{2} \tau^{2}}\right) r^{2} } ,
\end{equation}
where $\Omega = \rho/\rho_{c}$. Equation (\ref{eqn:phasespacederiv}) relates the velocity of the receding sources to their distance. Carmeli integrated (\ref{eqn:phasespacederiv}) and found the redshift distance relationship
\begin{equation} \label{eqn:rctau}
\frac {r} {c \tau}= \frac {\sinh \left(\beta \sqrt{1-\Omega}\right)} {\sqrt{1-\Omega}} ,  
\end{equation}
where, 
\begin{equation} \label{eqn:beta}
\beta = \frac{v}{c} = \frac{(1+z)^2-1}{(1+z)^2+1},
\end{equation}
and $\Omega$ the average matter density of the universe, a function of redshift $z$, must be found. 

This equation was used in both Hartnett \cite{Hartnett2005}, which forms the basis of the work presented here, and Oliveira and Hartnett \cite{Oliveira2006} where comparisons were made to the high redshift SNe Ia data from Riess \textit{et al} \cite{Riess2004}, Astier \textit{et al} \cite{Astier2005} and Knop \textit{et al} \cite{Knop2003}. The proper distance is converted to magnitude as follows.
\begin{equation} \label{eqn:lumindistance}
m(z) = \mathcal{M} + 5\,log \left[ \mathcal{D}_{L}(z;\Omega) \right], 
\end{equation}
where $\mathcal{D}_{L}$ is the dimensionless ``Hubble constant free'' luminosity distance. Refer \cite{Perlmutter1997,Riess1998}.
Here 
\begin{equation} \label{eqn:scriptem}
\mathcal{M} = 5\,log\left(\frac{c \tau}{Mpc}\right) + 25 + M_{B} + a.
\end{equation}
The units of $c \tau$ are $Mpc$. The constant $25$ results from the luminosity distance expressed in $Mpc$. However, $\mathcal{M}$ in (\ref{eqn:lumindistance}) represents a scale offset for the distance modulus (m-$M_{B}$). It is sufficient to treat it as a single constant chosen from the fit. In practice we use $a$, a small free parameter, to optimize the fits.

In CGR the luminosity distance was found \cite{Hartnett2007} to be slightly different to the expression used in the FRW theory. This fact, results from a relativistic effect on the emission times of the photons from the distant source.  In CGR, times at cosmological distances add according to a relativistic addition law, \cite{Carmeli2002ab} when referred to the observer at $t = 0$. When this is taken into account we get an additional factor $(1-\beta^2)^{-1/2}$, hence $\mathcal{D}_{L}$ is given by
\begin{equation} \label{eqn:magnitude}
\mathcal{D}_{L}(z;\Omega) = \frac {r} {c \tau} (1+z)\left(1-\beta^2 \right)^{-1/2}  
\end{equation}
using (\ref{eqn:rctau}), which is a function of $\Omega$ and $z$.

The parameter $\mathcal{M}$ incorporates the various parameters that are independent of the redshift, $z$. The parameter $M_{B}$ is the absolute magnitude of the supernova at the peak of its light-curve and the parameter $a$ allows for any uncompensated extinction or offset in the mean of absolute magnitudes or an arbitrary zero point. It also acts as a correction to the value of $\tau$ since $\mathcal{M}$ contains it. The absolute magnitude then acts as a ``standard candle'' from which the luminosity and hence distance can be estimated. 

The value of $M_{B}$ need not be known, neither any other component in $\mathcal{M}$, as $\mathcal{M}$ has the effect of merely shifting the fit curve (\ref{eqn:magnitude}) along the magnitude axis. However by choosing the value of the Hubble-Carmeli constant $\tau = 4.28 \times 10^{17} \, s = 13.58 \; Gyr$, which is the reciprocal of the chosen value of the Hubble constant in the gravity free limit $h = 72.17 \pm 0.84$ (statistical) $km.s^{-1} Mpc^{-1}$ \cite{Oliveira2006} $\mathcal{M} = 43.09 + M_{B} + a$.  

\section{\label{sec:Extended}Extended redshift range}

\begin{figure}
\includegraphics[width = 5 in]{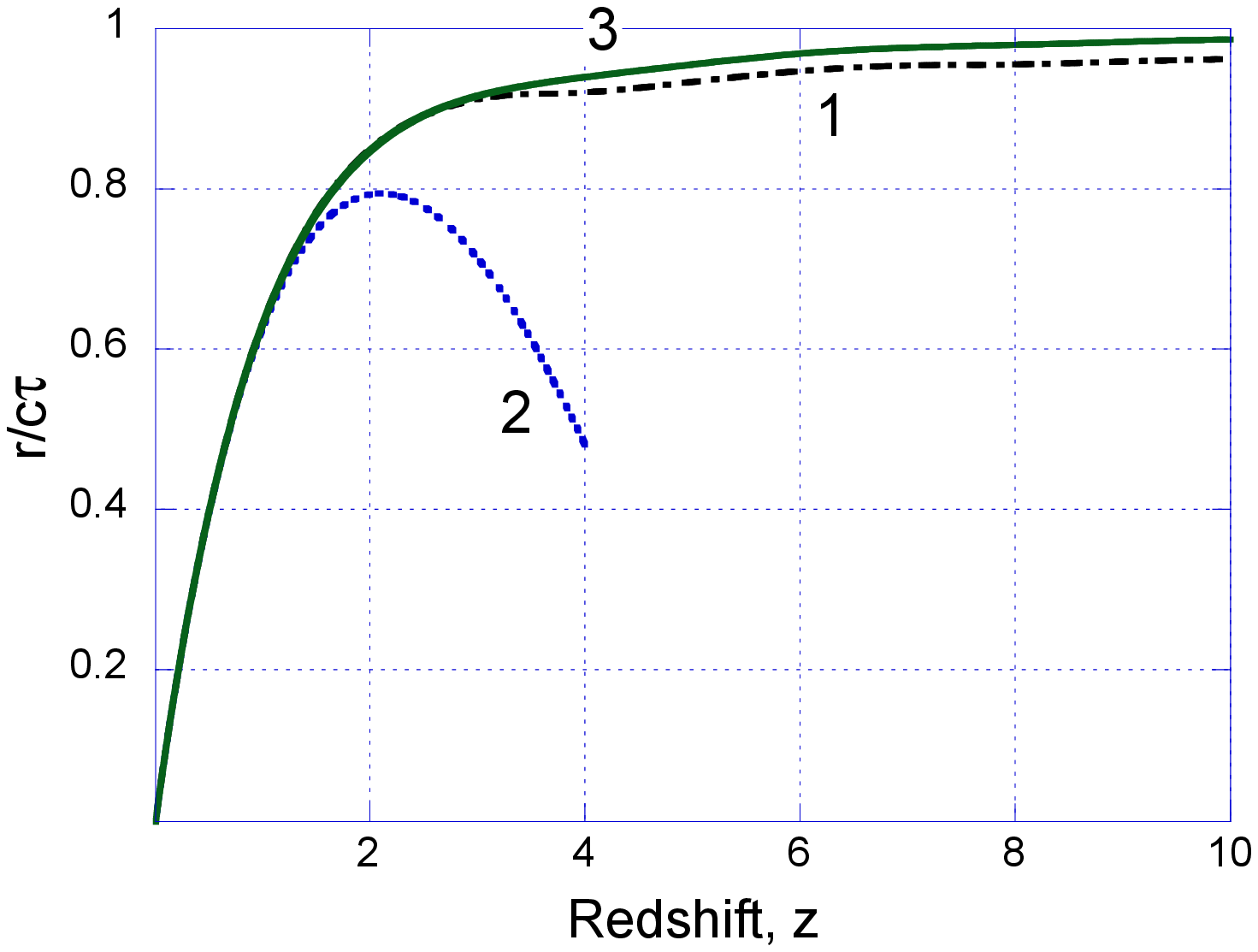}
\caption{\label{fig:fig1} Redshift distance relation: Curve 1 represents $r/c \tau$ from the solution of (\ref{eqn:drdz2}) with $\Omega(z) = \Omega_m (1+z)^3$,  curve 2 represents $r/c \tau$ from (\ref{eqn:rctau}) with $\Omega(z) = \Omega_m (1+z)^3$ and curve 3 represents $r/c \tau$ from the solution of (\ref{eqn:drdz2}) but with the density taken from curve 1 of fig. \ref{fig:fig2}.}
\end{figure}

The proper distance from (\ref{eqn:rctau}) however has limited application because it has essentially been found by integrating (\ref{eqn:phasespacederiv}) assuming constant density then using (\ref{eqn:beta}). The result being only valid over a limited redshift range. In order to improve on this, because the observables are  distance (magnitude) and redshift not velocity, the differential equation (\ref{eqn:phasespacederiv}) must first be converted to $dr/dz$ using the chain rule as follows.
\begin{equation} \label{eqn:drdz}
\frac{1}{\tau^2}\left(\frac{dr}{dz}\frac{dz}{dv} \right)^2=  1+ \frac{1-\Omega(z)}{c^{2} \tau^{2}} r^{2} ,
\end{equation}
where it must be remembered that matter density $\Omega$ is also a function of redshift. The derivative $dv/dz$ is then calculated from (\ref{eqn:beta}) and substituted into (\ref{eqn:drdz}), which becomes
\begin{equation} \label{eqn:drdz2}
\frac{1}{c^2 \tau^2}\left(\frac{dr}{dz} \right)^2= \frac{16 (1+z)^2}{(1+(1+z)^2)^4} \left(1+ (1-\Omega(z)) \frac{r^{2}}{c^{2} \tau^{2}}  \right).
\end{equation}
This differential equation is then solved for $r/c \tau$, but because it involves mixed terms on the rhs, it must be numerically solved. But we can employ a `bootstrapping' technique. By substituting the limited redshift solution from (\ref{eqn:rctau}) into the rhs and assuming the averaged matter density in the universe
\begin{equation} \label{eqn:density}
\Omega(z) = \Omega_m (1+z)^3,
\end{equation}
equation (\ref{eqn:drdz2}) was numerically solved for $r/c \tau$. The result is shown in curve 1 of fig. \ref{fig:fig1}. This is compared with $r/c \tau$ resulting from the initial equation (\ref{eqn:rctau}) represented by curve 2. 

Then $r/c \tau$ (of curve 1) was then taken as input on the rhs of (\ref{eqn:drdz2}) and iterated again to numerically solve for $r/c \tau$.  However because of the requirement (\ref{eqn:condition}) on the solution from the field equations the result is only valid to about $z = 2.5$ with the assumed density dependence (\ref{eqn:density}) on the present epoch density $\Omega_m = 0.04$. Dimensionally the matter density depends on the length cubed if we assume constant mass, but the precise density dependence is unknown at high redshifts. Therefore curve 1 of fig. \ref{fig:fig1} plotted beyond $z = 2.5$ can only be considered approximate. 

Now in CGR there is the scale radius 
\begin{equation} \label{eqn:scaleradius}
R_0 = \frac{c \tau}{\sqrt{|1-\Omega|}}.
\end{equation}
From (\ref{eqn:scaleradius}) and the condition (\ref{eqn:condition}) the range of validity of the solution (\ref{eqn:phasespacederiv}) is $0 \leq r/c \tau < 1 \leq R_0$ if  $\Omega \leq 1$ and $0 \leq r/c \tau < R_0$ if $\Omega > 1$. This means the solution is valid for all values of the matter density $\Omega(z) < 2$. 

\begin{figure}
\includegraphics[width = 5 in]{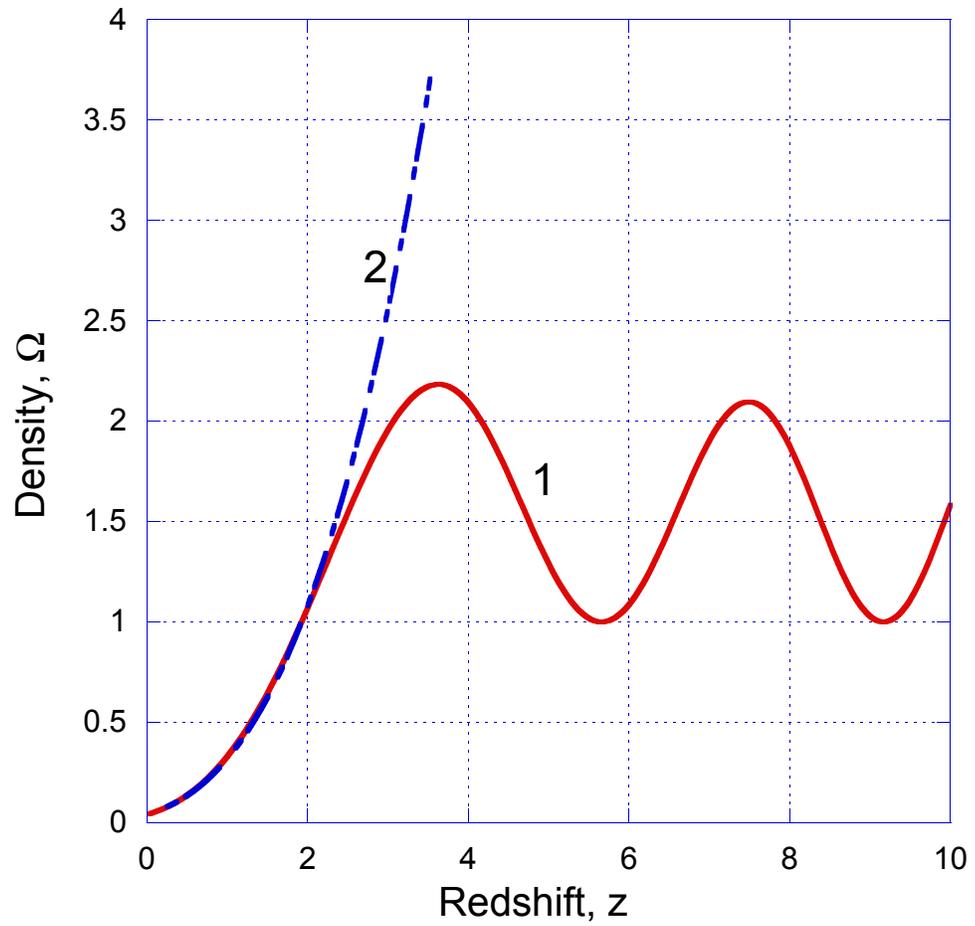}
\caption{\label{fig:fig2} Density as a function of redshift: Curve 1 represents density derived from (\ref{eqn:drdz2}) and curve 2 represents the density function $\Omega(z) = \Omega_m (1+z)^3$ with $\Omega_m = 0.04$.}
\end{figure}

Now if we again use the solution (curve 1 of fig. \ref{fig:fig1}) and  solve (\ref{eqn:drdz2}) for matter density $\Omega$, it results in a density function such that (\ref{eqn:condition}) is always true. The density function is shown as curve 1 in fig. \ref{fig:fig2} and is compared with the initial density function (\ref{eqn:density}) shown as curve 2. Finally using this as the density function in (\ref{eqn:drdz2}) and numerically solving for $r/c \tau$ results in curve 3 in fig. \ref{fig:fig1}.

\begin{figure}
\includegraphics[width = 5 in]{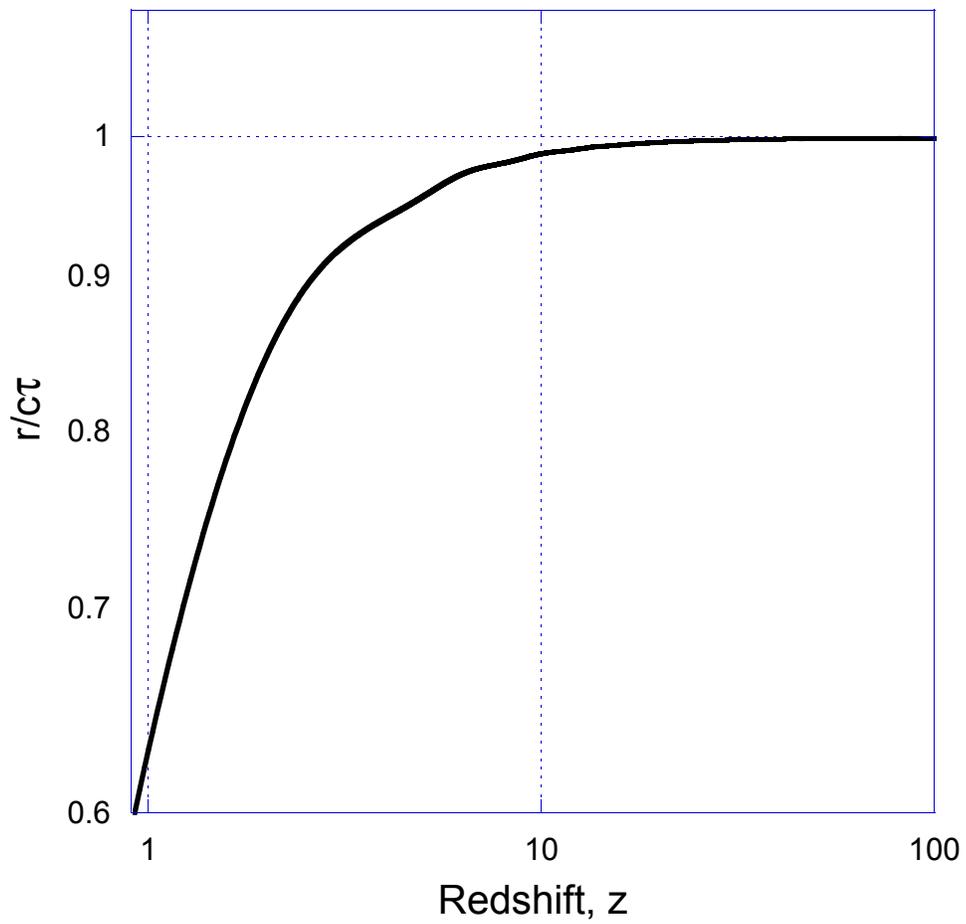}
\caption{\label{fig:fig3} Redshift distance relation to $z = 100$ using density function of curve 1 from fig. \ref{fig:fig2}.}
\end{figure}

The new density function has some unexpected features, but it produces a smooth monotonically increasing function of distance on redshift as expected. See fig. \ref{fig:fig3} where it has been plotted to $z = 100$ for the case where $\Omega_m = 0.04$. The fluctuating part of $\Omega(z)$ for $z > 4$ is approximately 
\begin{equation} \label{eqn:sinedensity}
1 + 1.1\; cos^2\left[\beta \sqrt{0.04(1+z)^3-1} +1.6 \right],
\end{equation}
where $\beta$ is determined from (\ref{eqn:beta}). The function (\ref{eqn:sinedensity}) has a decreasing period as a function of $z$, a minimum at $\Omega = 1$ and a maximum slightly greater than $\Omega = 2$. Any density function that remains within the range of this function for $z > 4$, even a constant density, will yield a smooth monotonic redshift-distance relation. However the oscillating density as a function of redshift could represent episodic creation from the vacuum as the universe expands. Particle production at those epochs is a byproduct of the expansion process. Gemelli \cite{Gemelli} found from a hydrodynamic solution of the 5D problem that particle production must occur from the expansion provided the Universe is not isentropic.

\section{\label{sec:Fits}Quality of curve fits}

\begin{figure}
\includegraphics[width = 5 in]{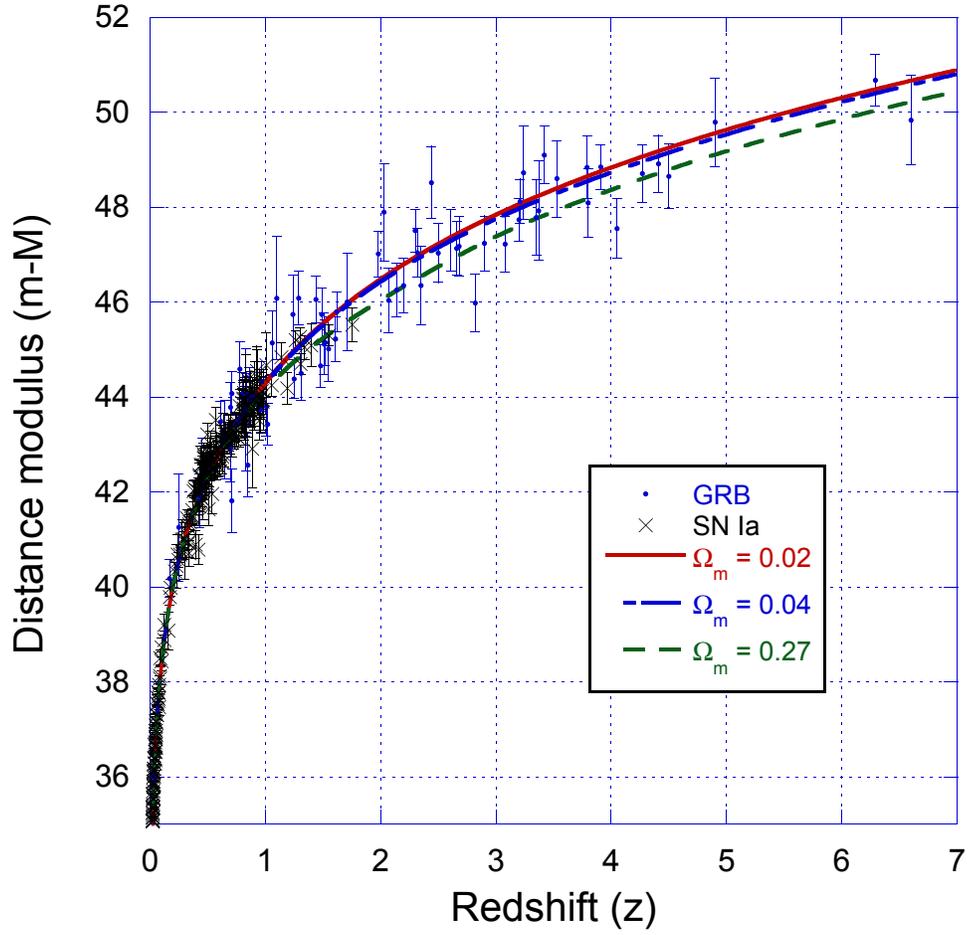}
\caption{\label{fig:fig4} Distance modulus data from the combined SN Ia data set of Riess \textit{et al} and Astier \textit{et al} (crosses) and the GRB data (dots) derived from Schaefer \cite{Schaefer2007} with curve fits using the extended equation (\ref{eqn:drdz2}). The top solid curve represents the distance modulus where $\Omega_m = 0.02$, the broken curve below it is where $\Omega_m = 0.04$ and the dashed curve below the latter is where $\Omega_m = 0.27$.}
\end{figure}

The initial equation (\ref{eqn:rctau}) has been curve fitted to SNe Ia data \cite{Oliveira2006} as described in section \ref{sec:Comparison} and excellent fits resulted without any dark matter component. The best statistical fit resulted in a value of $\Omega_m = 0.0401 \pm 0.0199$, which is consistent with the observed baryonic matter density \cite{Fukugita1998}. The fit was accomplished with the combined Gold and Silver SNe Ia data sets of Riess \textit{et al} \cite{Riess2004} with that of Astier \textit{et al} \cite{Astier2005} that extended to $z=1.75$, the current limit of observational data. Inspection of the $r/c\tau$ curve (curves 1 or 3 shown in fig. \ref{fig:fig1}) and the initial curve (curve 2)  for the region $1.5 < z < 1.75$ indicates that only the last two SNe Ia data would be sensitive to the numerically determined extension here. See crosses in fig. \ref{fig:fig4}. 

Using $r/c\tau$ from the extended equation (\ref{eqn:drdz2}) I have curve fitted to the data set of Riess \textit{et al} and also to combined data set of Riess \textit{et al} and Astier \textit{et al}. However before using Astier \textit{et al} data I added $0.14$ magnitudes to their quoted errors to align their errors with those of Riess \textit{et al}. Otherwise Astier \textit{et al} seem to have underestimated their magnitude errors, which unfairly bias the $\chi^2$ calculations. The resulting $\chi^2$s for the model here are shown in Table I.
 
Clearly more SN Ia data, at higher redshifts, are needed to test this prediction. Consequently GRBs have been used to get distance modulus and redshift as independent parameters (from Schaefer \cite{Schaefer2007}) and are shown as dots in fig. \ref{fig:fig4}. However the magnitude calculation is not really independent of the chosen cosmology as explained in the latter reference. Nevertheless using the data from columns 2 and 8 of Table 6 \cite{Schaefer2007}, and making the necessary changes to convert the magnitudes from the FRW Concordance Model to that of the Carmeli model, new distance modulii are obtained. 

\begin{figure}
\includegraphics[width = 5 in]{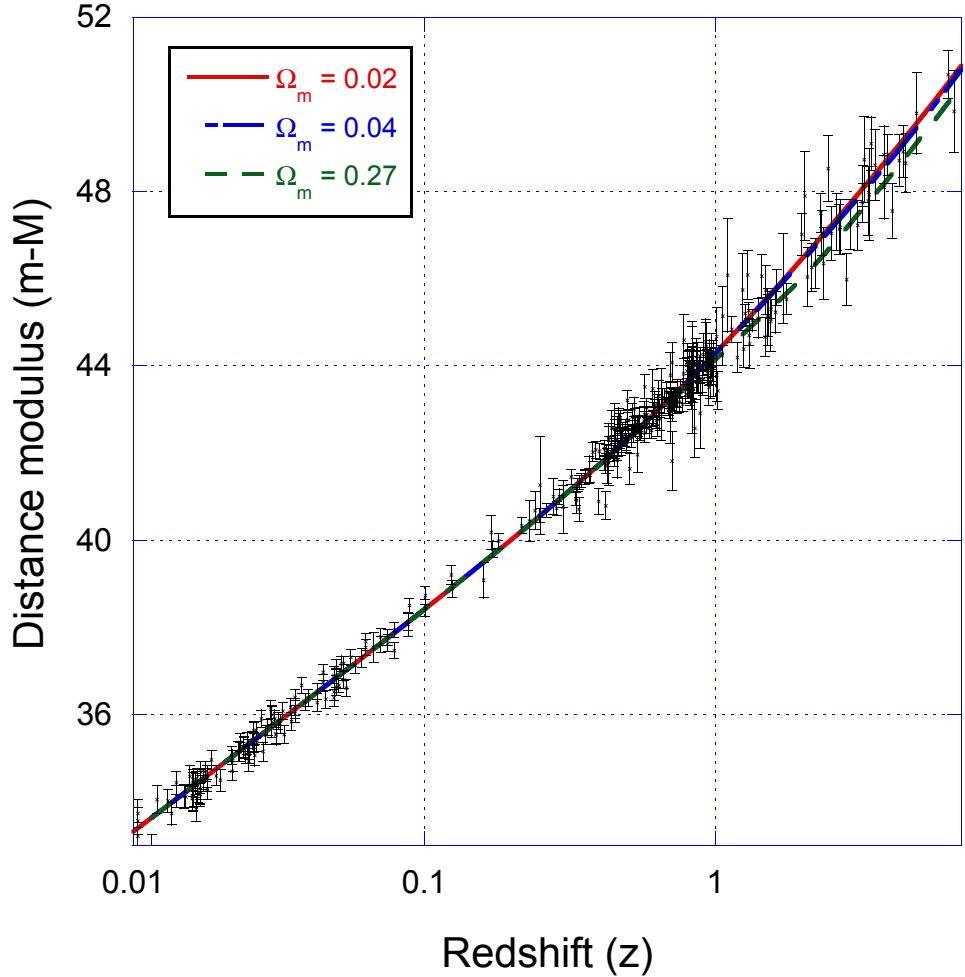}
\caption{\label{fig:fig5} All 371 distance modulus data from the combined SN Ia data set of Riess \textit{et al} and Astier \textit{et al} and the GRB data derived from Schaefer \cite{Schaefer2007} with curve fits as shown in fig. \ref{fig:fig4}, but on a redshift log scale. The top solid curve represents the distance modulus where $\Omega_m = 0.02$, the broken curve below it is where $\Omega_m = 0.04$ and the dashed curve below the latter is where $\Omega_m = 0.27$.}
\end{figure}

\begin{figure}
\includegraphics[width = 5 in]{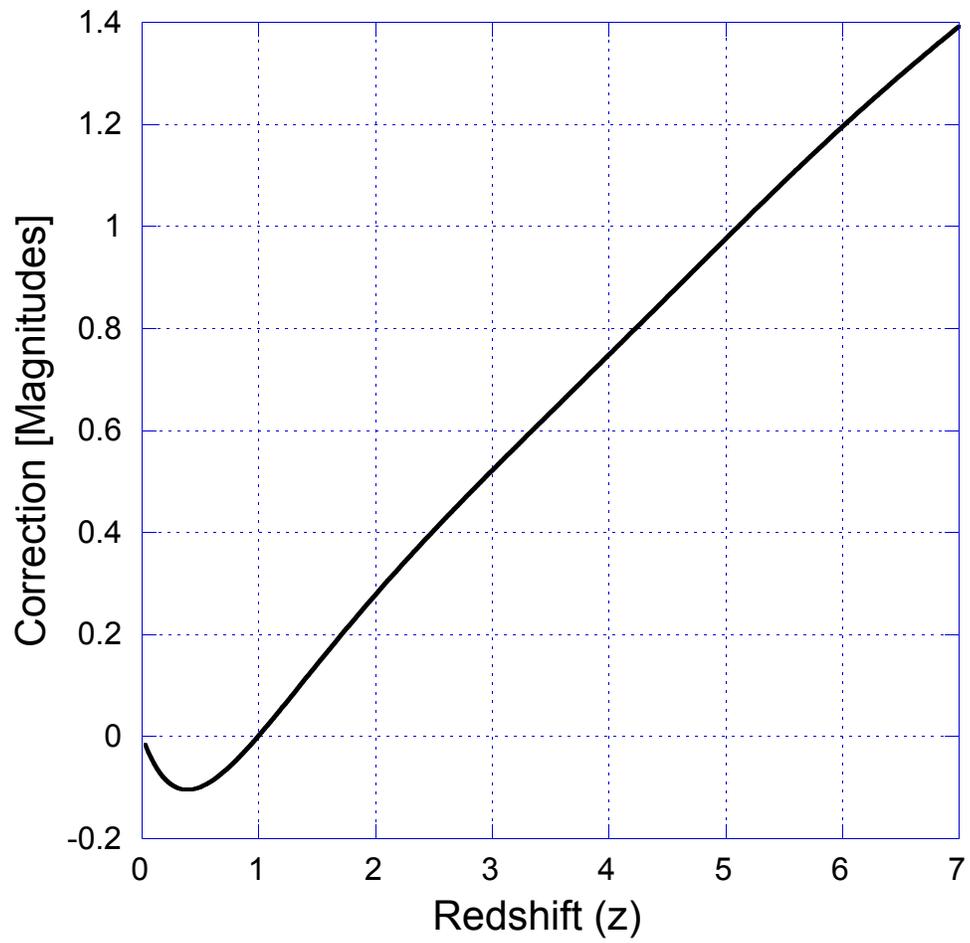}
\caption{\label{fig:fig6} The correction applied to the distance modulus data of Schaefer \cite{Schaefer2007} to account for the Carmeli cosmology.}
\end{figure}

In order to use the data from Schaefer \cite{Schaefer2007} first the specific details particular to a cosmological model he used are subtracted then the specific details from the CGR theory are applied. This means small corrections are made and are shown in fig. \ref{fig:fig6} as a function of redshift. This also means the magnitudes calculated this way are not really independent of our chosen CGR cosmology either. In fact, problems related to the cosmological dependence of the GRBs calibration, in principle, may invalidate such a procedure. Nevertheless we may get some benefit from this exercise. In figs \ref{fig:fig4} and \ref{fig:fig5} both SN Ia and GRB data are represented on the same plot. However the SN Ia magnitudes are not dependent on any particular cosmology and present a much stronger constraint on the applied theory. Because it is shape of the curve fitting to the data that carries the most information if the theory fits the SN Ia data it should also fit the GRB data.

The following are the corrections applied to convert the distance modulus of Schaefer to that of the extended theory of CGR developed here. Since the luminosity distance in CGR is different to that in FRW theory, by the factor $(1-\beta^2)^{-1/2}$, it's effect must be added to the distance modulus. Therefore the luminosity distance equation (10) of Schaefer must be replaced by that in CGR. So equation (\ref{eqn:magnitude}), with $r/c\tau$ determined from the numerical solution of (\ref{eqn:drdz2}) and  $\Omega_m = 0.04$, was used to calculate a correction, which was added as magnitude to the distance modulus. Equation (10) of Schaefer \cite{Schaefer2007} with $\Omega_m = 0.27$ and $\Omega_{\Lambda} = 0.73$ for the Concordance Model was numerically integrated and removed as a magnitude from the data. These three corrections have the combined effect on the data as shown in fig. \ref{fig:fig6}.

Finally by fitting the extended equation (\ref{eqn:drdz2}) with $\Omega_m = 0.04$ to this data and to the combined set of Riess \textit{et al} and Astier \textit{et al}  it was determined that $0.1988$ magnitudes needed to be added to shift the GRB set so it could be represented on the same plot as the SN Ia data. This was done by determining the best fit value of $a$ from (\ref{eqn:scriptem}) for each data set then adding the difference to the GRB distance modulii so $a$ is the same for both. Hence a fit to the total data set can then be accomplished. See fig. \ref{fig:fig5}. There the data and fits using the extended Carmeli theory with $\Omega_m = 0.02, 0.04$ and $0.27$ are shown on a redshift log axis. This better shows the low redshift fit. At high redshifts the fits with the smaller values of $\Omega_m$ are favored.

In order quantify the goodness of the least squares fitting I have used the $\chi^{2}$ parameter which measures the goodness of the fit between the data and the theoretical curve assuming the two fit parameters $a$ and $\Omega_m$. Hence $\chi^{2}$ is calculated from
\begin{equation} \label{eqn:chisq}
\chi^2 = \sum_{i=1}^{N} \frac{1}{\sigma_i^2} \left[(m-M)(z)_i-(m-M)(z_{obs})_i\right]^2,
\end{equation}
where $N$ are the number of data; $(m-M)(z)$ are determined from (\ref{eqn:lumindistance}) with fit values of $a$ and $\Omega_m$; $(m-M)(z_{obs})$ are the observed distance modulus data at measured redshifts $z_{obs}$; $\sigma_i$ are the published magnitude errors. The values of $\chi^2/N$ ($\approx \chi_{d.o.f}^2$) are shown in Table I, calculated using published errors on the distance modulus data. The best fit value of $a$ is found at low redshift and then applied to all fits with various values of $\Omega_m$.

\begin{table}[ph]
\center
\small
Table I:~Curve fit $\chi^2$ parameter\\
\vspace{6pt}
\begin{tabular}{c|cc} \hline \hline
Data set(s) 										&$\Omega_m$& $\chi^2$/N \\
\hline
SNe Ia 													& 0.00  & 1.4120 \\
Riess et al Gold \& Silver 			& 0.02	& 1.4047 \\
$N$ = 185 												& 0.04 	& 1.3987 \\
$a$ = 0.2400											& 0.06 	& 1.3941 \\
($a$ from $z < 0.23$)								& 0.10	& 1.3888 \\
 																& 0.27 	& 1.4258 \\
\hline
SNe Ia														& 0.00  & 1.2366 \\
Riess et al + Astier et al SNLS & 0.02	& 1.2341 \\
$N$ = 302 												& 0.04 	& 1.2326 \\
$a$ =  0.1894											& 0.06 	& 1.2321 \\
($a$ from $z < 0.24$)								& 0.10	& 1.2340 \\
 																& 0.27 	& 1.2871 \\
\hline
SNe Ia  + GRBs 										& 0.00  & 1.2191 \\
Riess + Astier + Schaefer 			& 0.02	& 1.2005 \\
$N$ = 371	 												& 0.04 	& 1.1902 \\
$a$ = 0.2113											& 0.06 	& 1.1860 \\
($a$ from $z < 2$)									& 0.10	& 1.1870 \\
 																& 0.27 	& 1.2538 \\
 				
\end{tabular}
\end{table}

From Table I the smallest $\chi^2$/N value are generally obtained where $0.04 < \Omega_m < 0.06$, especially when the GRBs are included. But the quality of the fits are not strongly dependent on the matter density. The locally measured baryonic matter budget has determined that $0.007 < \Omega_b < 0.041$ \cite{Fukugita1998} where a Hubble constant of $70 \; km.s^{-1} Mpc^{-1}$ was assumed. This region is consistent with the most probable values of $\Omega_m$ from this analysis. Definitely a matter density of $\Omega_m = 0.27$ is not necessary for the best fit. Therefore no exotic dark matter need be assumed. However to properly test this model much more SNe Ia data are needed where $z > 2$ and much better constrained magnitude errors. The GRB data are not determined independently of the cosmology tested and therefore by themselves don't provide a `standard candle' test of the theory.

\section{\label{sec:flatspace}Spatially flat universe}

In section \ref{sec:phasespace} the time part of the metric $g_{\mu \nu}$ was ignored. Here we use coordinates  $x^{\mu}=(ct, r, \theta, \phi, \tau v)$ where $\mu = 0-4$, but the full 5D description of cosmos has never been determined in the Carmelian theory. So the question must be answered, ``What is the $g_{00}$ metric component for the large scale structure of the Universe in CGR?''

First note from (\ref{eqn:soln1field2}) and (\ref{eqn:soln1field3}) the $g_{11}$ metric component 
\begin{equation} \label{eqn:g11}
g_{11} = -\left(1+\frac{1-\Omega}{c^2\tau^2}r^2\right)^{-1}.
\end{equation}
Using the scale radius we can define an energy density from the curvature
\begin{equation} \label{eqn:curvaturedensity}
\Omega_K = \frac{c^2}{h^2 R_0^2} = \frac{c^2 \tau^2}{R_0^2},
\end{equation}
which, when we use (\ref{eqn:scaleradius}), becomes 
\begin{equation} \label{eqn:curvaturedensity2}
\Omega_K = 1-\Omega.
\end{equation}
This quantifies the energy in the curved \textit{spacevelocity}.

In the FRW theory the energy density of the cosmological constant is defined $\rho_{\Lambda} = \Lambda/8 \pi G$ hence
\begin{equation} \label{eqn:lambdadensityFRW}
\Omega_{\Lambda} = \frac{\Lambda}{3 H^2_0}.
\end{equation}
Even though the cosmological constant is not explicitly used in CGR, it follows from the definition of the critical density that
\begin{equation} \label{eqn:criticaldensityCGR}
\rho_c = \frac{3}{8 \pi G \tau^2} = \frac{\Lambda}{8 \pi G}, 
\end{equation}
when the cosmological constant $\Lambda$ is identified with $3/\tau^2$. Therefore in CGR it follows that
\begin{equation} \label{eqn:lambdadensityCGR}
\Omega_{\Lambda} = \frac{\Lambda}{3 h^2} = \Lambda\left(\frac{\tau^2}{3}\right) = 1.
\end{equation}
In CGR $h = \tau^{-1}$. This means that in CGR the vacuum energy $\rho_{vac} = \Lambda/8\pi G$ is encoded in the metric via the critical density since $\rho_{eff} = \rho - \rho_c$ principally defines the physics. So $\Omega_{\Lambda} = 1$ identically and at all epochs of time. (The determination of $\Omega_{\Lambda}$ in \cite{Hartnett2005} was flawed due to an incorrect definition.) Also we can relate $\Omega_{\Lambda}$ to the curvature density by
\begin{equation} \label{eqn:curvaturedadensity3}
\Omega_K = \Omega_{\Lambda} - \Omega,
\end{equation}
which becomes
\begin{equation} \label{eqn:curvaturedadensity4}
\Omega_k = \Omega_{\Lambda} - \Omega_m,
\end{equation}
at the present epoch ($z \approx 0$). Here $\Omega = \Omega_m(1+z)^3$ ($z \ll 1$) and hence $\Omega_K \rightarrow \Omega_k$ as $z \rightarrow 0$.

Finally we can write for the total energy density, the sum of the matter density and the curvature density,
\begin{equation} \label{eqn:totaldensity}
\Omega_t = \Omega +  \Omega_K = \Omega +  1 - \Omega = 1,
\end{equation}
which means the present epoch value is trivially
\begin{equation} \label{eqn:totaldensity0}
\Omega_0 = \Omega_m +  \Omega_k = \Omega_m +  1 - \Omega_m = 1.
\end{equation}
This means that the 3D spatial part of the Universe is always flat as it expands. This explains why we live in a universe that we observe to be identically geometrically spatially flat. The curvature is due to the velocity dimension. Only at some past epoch, in a radiation dominated universe, would the total mass/energy density depart from unity. 

I therefore write
\begin{equation} \label{eqn:g00}
g_{00}(r) = 1 + (1-\Omega_t) r^2,
\end{equation}
where $r$ is expressed in units of $c \tau$. Equation (\ref{eqn:g00}) follows from $g_{00} = 1-4\Phi/c^2$  where $\Phi$ is taken from the gravitational potential but with effective density, which in turn involves the total energy density because we are now considering \textit{spacetime}. The factor 4 result from a comparison of the CGR theory with Newtonian theory.

Clearly from (\ref{eqn:totaldensity}) it follows that $g_{00}(r) = 1$ regardless of epoch.
Thus from the usual relativistic expression 
\begin{equation} \label{eqn:gravredshift}
1 + z_{grav} = \sqrt{\frac{g_{00}(0)}{g_{00}(r)}} =1,
\end{equation}
and the gravitational redshift is zero regardless of epoch. As expected if the emission and reception of a photon both occur in flat space then we'd expect no gravitational effects.

Therefore we can write down the full 5D line element for CGR in any dynamic spherically symmetrical isotropic universe,
\begin{equation} \label{eqn:linelement5D}
ds^2 = c^2dt^2 -\left(1+\frac{1-\Omega}{c^2\tau^2}r^2\right)^{-1}dr^2 +\tau^2 dv^2.
\end{equation}
The $\theta$ and $\phi$ coordinates do not appear due to the isotropy condition $d\theta = d\phi=0$. Due to the Hubble law the 2nd and 3rd terms sum to zero leaving $dt = ds/c$, the proper time. Clocks, co-moving with the galaxies in the Hubble expansion, would measure the same proper time.

However inside the Galaxy we expect the matter density to be much higher than critical, ie $\Omega_{galaxy} \gg 1$ and the total mass/energy density can be written
\begin{equation} \label{eqn:totaldensitygal}
\Omega_0 |_{galaxy} = \Omega_{galaxy} +  \Omega_k \approx \Omega_{galaxy},
\end{equation}
because $\Omega_k \approx 1$, since it is cosmologically determined. Therefore this explains why the galaxy matter density only is appropriate when considering the Poisson equation for galaxies.\cite{Hartnett2005b} 

As a result inside a galaxy we can write 
\begin{equation} \label{eqn:g00Galaxy}
g_{00}(r) = 1 + \Omega_K \frac{r^2}{c^2\tau^2} + \Omega_{galaxy} \frac{r^2}{c^2\tau^2} ,
\end{equation}
in terms of densities at some past epoch. Depending on the mass density of the galaxy, or cluster of galaxies, the value of $g_{00}$ here changes. As we approach larger and larger structures it mass density approaches that of the Universe as a whole and $g_{00} \rightarrow 1$ as we approach the largest scales of the Universe. Galaxies in the cosmos then act only as local perturbations but have no effect on $\Omega_K$. That depends only on the average mass density of the whole Universe, which depends on epoch ($z$).

Equation (\ref{eqn:g00Galaxy}) is in essence the same expression used on page 173 of Carmeli \cite{Carmeli2002a} in his gravitational redshift formula rewritten here as
\begin{equation} \label{eqn:Cgravz}
\frac{\lambda_2}{\lambda_1} = \sqrt{\frac{1+ \Omega_K r^2_2/c^2\tau^2-R_S/r_2}{1+\Omega_K r^2_1/c^2\tau^2-R_S/r_1}}.
\end{equation}
involving a cosmological contribution ($\Omega_K r^2/c^2\tau^2$) and $R_S=2GM/c^2$, a local contribution where the mass $M$ is that of a compact object. The curvature ($\Omega_K$) results from the averaged mass/energy density of the whole cosmos, which determines how the galaxies `move' but motions of particles within galaxies is dominated by the mass  of the galaxy and the masses of the compact objects within. Therefore when considering the gravitational redshifts due to compact objects we can neglect any cosmological effects, only the usual Schwarzschild radius of the object need be considered. The cosmological contributions in (\ref{eqn:Cgravz}) are generally negligible. This then leads back to the realm of general relativity.
  
\section{\label{sec:conclusion}Conclusion}
The solution of the Einstein field equation using the \textit{spacevelocity} of Moshe Carmeli requires the condition (\ref{eqn:condition}). This means that the redshift-distance relation can only be valid, in general, where $\Omega \leq 2$, in a matter dominated universe. If follows then if the Universe underwent periodic particle creation at high redshift, while the expansion progressed, then we can construct a Hubble Diagram that is valid over arbitrary redshift. This approach extends the Carmeli theory to where it makes a clear prediction and can be tested with future space-based telescopes, like the James Webb. 

The current plethora of cosmologies seem to require a few extra degrees of freedom, for example, scalar fields, dark energy or dark matter, but are they really capable of being distinguished one from another when tested against observations? If a future space-based telescope was able to resolve well enough type Ia supernovae to redshifts approaching  z = 10, then these various cosmologies, which fit the same data sets at much smaller redshifts, may be distinguished one from another.  I believe only at these very high redshifts, provided the uncertainties on the data are small enough, can we hope to accomplish this task and really test Carmeli's cosmology.

However using the Carmelian metric we are able to answer the question, `Why do we observe the Universe at this present epoch to be Euclidean?'  Carmeli's choice of metric implicitly involves a vacuum energy density, such that when the matter density of the Universe reaches that of the vacuum density the fabric of space is relaxed and the expansion coasts (i.e. is unaccelerated). The requirement on the energy content of the vacuum leads to the Universe being spatially flat at all epochs.  This is because the total curvature of the Universe ($\Omega_k=1-\Omega_m$ at the present epoch) is the result of the average matter content warping \textit{spacevelocity} away from flatness whereas the 3D spatial part remains flat. So at the present epoch $\Omega_0 = \Omega_k + \Omega_m = 1$.  

\section{Acknowledgment}
I would like to sincerely acknowledge the encouragement and support of the late Prof. Moshe Carmeli of Ben Gurion University, Beer Sheva, Israel who passed away on September 26th 2007. My condolences and prayers go out to his wife Elishava Carmeli and family. This work was supported by the Australian Research Council

\newpage

\end{document}